\documentclass{elsarticle}

\usepackage{amssymb}
\usepackage{xcolor}
\usepackage{tikz-cd}
\usepackage{floatrow}
\usepackage{caption}
\usepackage[section]{placeins}

\newcommand{\myclosearrowright}{\mathrel{\mathrm{\tikz[baseline=-0.5ex] \draw (0,0) (0em,0.3ex)--(1.5em,0.3ex)--(1.5em,0.7ex)--(2em,0ex)--(1.5em,-0.7ex)--(1.5em,-0.3ex)--(0em,-0.3ex)--cycle;}}}
\newcommand{\myclosearrowleft}{\mathrel{\mathrm{\tikz[baseline=-0.5ex] \draw (0,0) (0em,0.3ex)--(-1.5em,0.3ex)--(-1.5em,0.7ex)--(-2em,0ex)--(-1.5em,-0.7ex)--(-1.5em,-0.3ex)--(0em,-0.3ex)--cycle;}}}

\newcommand{\myarrowleft[1]}{\tikz[baseline=-0.5ex]\draw[color=#1] (0,0) (0em,0ex)--(-1.5em,0ex) (-1.2em,0.3ex)--(-1.5em,0ex)--(-1.2em,-0.3ex);}

\newcommand{\myoplus[1]}{\tikz[baseline=-0.5ex]\draw[color=#1] (0,0) circle (0.75ex) (-0.6ex,0)--(0.6ex,0) (0,0.6ex)--(0,-0.6ex);}

\newcommand{\myominus[1]}{\tikz[baseline=-0.5ex]\draw[color=#1] (0,0) circle (0.75ex) (-0.6ex,0)--(0.6ex,0);}

\makeatletter
\def\ps@pprintTitle{%
 \let\@oddhead\@empty
 \let\@evenhead\@empty
 \def\@oddfoot{}%
 \let\@evenfoot\@oddfoot}
\makeatother

\begin{document}

\title{Coulomb drag and counterflow Seebeck coefficient in bilayer-graphene double layers}

\author[pece,pbnc,nist,jqi]{J.~Hu\corref{cor}}
\ead{jiuning.hu@nist.gov}

\author[pphys,pbnc]{T.~Wu}
\author[pphys,pbnc]{J.~Tian}
\author[nist,jqi]{N.N.~Klimov}
\author[nist]{D.B.~Newell}

\author[pphys,pbnc,pece]{Y.P.~Chen\corref{cor}}
\ead{yongchen@purdue.edu}

\cortext[cor]{Corresponding authors}

\address[pece]{School of Electrical and Computer Engineering, Purdue University, West Lafayette, Indiana 47907, USA}
\address[pbnc]{Birck Nanotechnology Center, Purdue University, West Lafayette, Indiana 47907, USA}
\address[pphys]{Department of Physics and Astronomy, Purdue University, West Lafayette, Indiana 47907, USA}
\address[nist]{Physical Measurement Laboratory, National Inistitute of Standards and Technology, Gaithersburg, Maryland 20899, USA}
\address[jqi]{Joint Quantum Institute, University of Maryland, College Park, MD 20742, USA}

\begin{abstract}
We have fabricated bilayer-graphene double layers separated by a thin ($\sim$20 nm) boron nitride layer and performed Coulomb drag and counterflow thermoelectric transport measurements. The measured Coulomb drag resistivity is nearly three orders smaller in magnitude than the intralayer resistivities. The counterflow Seebeck coefficient is found to be well approximated by the difference between Seebeck coefficients of individual layers and exhibit a peak in the regime where two layers have opposite sign of charge carriers. The measured maximum counterflow power factor is $\sim$ 700 $\mu$W/K$^2$cm at room temperature, promising high power output per mass for lightweight thermoelectric applications. Our devices open a possibility for exploring the novel regime of thermoelectrics with tunable interactions between n-type and p-type channels based on graphene and other two-dimensional materials and their heterostructures.
\end{abstract}


\maketitle

\section{Introduction}
Graphene and boron nitride (BN) based heterostructures have attracted intense attention in recent years.\cite{dean2010boron,taychatanapat2011quantum,dean_multicomponent_2011,britnell2012electron,geim2013van} With the flake transfer technique\cite{jiao_creation_2008,jiao_selective_2009,dean2010boron,taychatanapat2011quantum} that enables convenient stacking of a large variety of two-dimensional layered materials, high quality samples of graphene/BN/graphene (where the BN thickness can be as thin as down to $\sim$1 nm\cite{gorbachev2012strong}) and other heterostructures have been fabricated to study the interlayer interactions. For example, strong Coulomb drag\cite{gorbachev2012strong,lee2016giant,li2016negative} and tunable metal-insulator transition\cite{ponomarenko2011tunable} have been observed. Vertical field-effect transistor\cite{georgiou2013vertical} and resonant tunneling\cite{britnell2013resonant,mishchenko2014twist} using atomically thin barriers have been demonstrated, promising for applications in logic and high-frequency devices. The electron-hole symmetry allows each layer of graphene to be populated with electrons or holes using gates, which is usually difficult to achieve in traditional semiconductor quantum wells.\cite{manfra2014molecular} In the regime where one graphene layer is p-type while the other is n-type, the intriguing exciton condensation\cite{eisenstein2004bose} has been predicted, though the predicted transition temperature spans a wide range.\cite{min2008room,lozovik2008electron,kharitonov2008electron,zhang2008excitonic,lozovik2012condensation,abergel2012density,sodemann2012interaction,suprunenko2012phases,perali2013high,abergel2013interlayer} A recent proposal suggests that in bilayer-graphene (BLG) double layers, the transition temperature could be well above liquid helium temperatures at zero magnetic field.\cite{perali2013high} This novel phase transition can be detected by performing counterflow transport (with equal magnitude and opposite direction of current in two layers), Coulomb drag and other measurements.\cite{eisenstein2004bose,nandi2012exciton} An anomalous negative Coulomb drag is observed in BLG double layers,\cite{lee2016giant,li2016negative} which is different from the drag in monolayer graphene double layers.\cite{kim2012coulomb,gorbachev2012strong} Signatures of exciton superfluid have been recently observed in BLG in magnetic field with much higher transition temperature than that in double quantum wells.\cite{liu2016quantum,li2016excitonic,eisenstein2004bose}
In addition, thermoelectric transport is a powerful tool for studying not only the single particle transport,\cite{goupil2011thermodynamics} but also strongly correlated systems\cite{rowe2005thermoelectrics,lin2014good} and macroscopic quantum coherence\cite{clarke1972superconducting,karpiuk2012superfluid,ranccon2014bosonic} such as exciton condensation. Thermoelectric transport might also be relevant to understanding the drag measurements in BLG double layers.\cite{lee2016giant} Nevertheless, no counterflow thermoelectric transport measurements have been performed.

As another motivation for the current work, high efficiency thermoelectric modules have been pursued for decades to improve solid-state thermoelectric generators and Peltier refrigerators. Due to often conflicting parameters, producing such modules requires careful material and structural engineering.\cite{snyder2008complex} High thermoelectric efficiency in low dimensional and nano-materials has been demonstrated, particularly for ``electron-crystal-phonon-glass" systems that have high electrical conductivity and low thermal conductivity.\cite{dresselhaus2007new,snyder2008complex} These traditional approaches focus on engineering the properties of a given material to enhance the thermoelectric figure of merit $ZT=\frac{\sigma S^2T}{\kappa}$, where $\sigma$ is electrical conductivity, $S$ is Seebeck coefficient, $\kappa$ is thermal conductivity and $T$ is temperature. Traditional thermoelectric units consist of two spatially separated and oppositely doped (p-type and n-type) semiconductor channel materials. The typical size of the channel materials and their separation are macroscopic, and their mutual interaction is negligible. In nanoscale thermoelectric devices, if the two channels are brought sufficiently close together, in principle the Coulomb interaction between the charge carriers in the two channels can become notable. It remains unknown how such interaction may affect the thermoelectric transport properties. Such a previous unexplored regime in thermoelectric devices may be studied in closely separated double layer systems, such as graphene/BN/graphene. 

In this paper, Coulomb drag resistivity (which arises from the interlayer interaction) and counterflow Seebeck coefficient are measured in two BLG layers separated by a thin BN layer shown in Figure \ref{fig:f1-2}. The Coulomb drag resistivity and Seebeck coefficient in each BLG layer show the expected sign when the carrier type in each layer is changed. In our current devices, the counterflow Seebeck coefficient can be well approximated by the sum of Seebeck coefficients from the individual layers, suggesting that the effect of the interlayer interaction on the counterflow thermoelectric transport is negligibly small (although the interlayer interaction is sufficient to give a measurable Coulomb drag signal). The magnitude of counterflow Seebeck coefficient and the calculated counterflow power factor increase with temperature. We provide a quantitative analysis about the thermoelectric performance in the counterflow regime. The maximum power factor at room temperature can be $\sim$ 700 $\mu$W/K$^2$cm, exceeding that of the good thermoelectric material Bi$_2$Te$_3$ at least by a factor of 5.\cite{ovsyannikov2008giant} Such a high power factor could be useful in thermoelectric applications, even though $ZT$ value is small due to high thermal conductivity of graphene. The possible impact of Coulomb drag on the counterflow Seebeck coefficient, though small in our samples, can be enhanced in devices with smaller separation between the two channels and could open new possibilities in developing high $ZT$ thermoelectric materials and structures.

\section{Methods}

\begin{figure}
	\includegraphics[width=10cm]{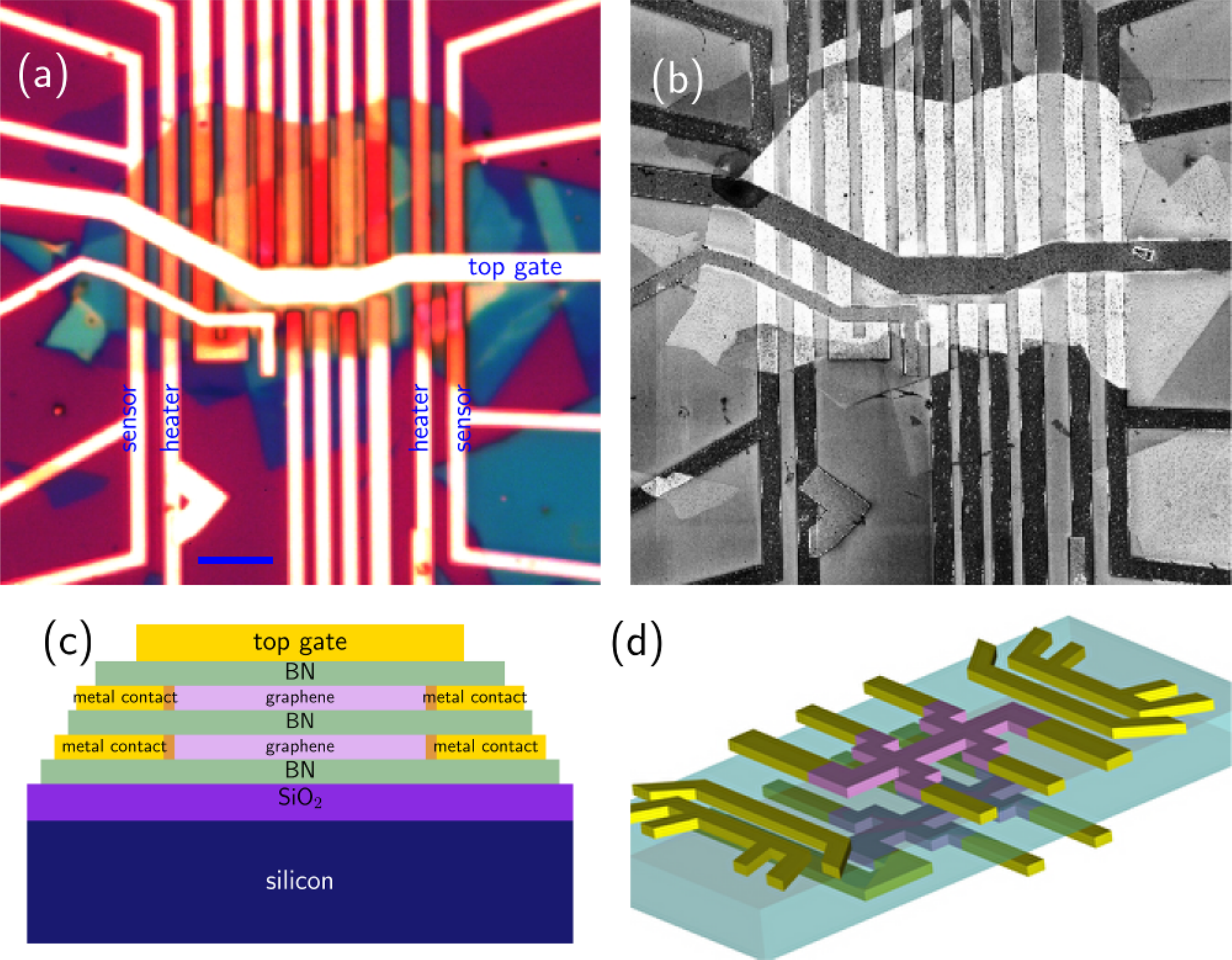}
	\caption{Optical image (a) of the graphene device (heater lines, temperature sensors and top gate are labeled in a), the corresponding SEM image (b), the schematics of cross section (c, heaters and sensors are not shown) and the three dimensional schematics of the graphene/BN/graphene stacking with metal contacts, heaters and sensor lines (purple for graphene, yellow for metal and transparent cyan for BN). The length of the scale bar in (a) is 5 $\mu$m.}
	\label{fig:f1-2}
\end{figure}

The measured graphene device in Figure \ref{fig:f1-2}a-b consists of two BLG layers separated by a thin BN layer ($\sim$20 nm thick). They have isolated metal contacts, as shown by the schematics in Figure \ref{fig:f1-2}c-d. The bilayer nature of graphene is confirmed by Raman spectroscopy and quantum Hall measurement. This layered structure is sitting on a local BN dielectric on SiO$_2$/silicon substrate and is covered by the top gate BN dielectric and metal. To fabricate the sample, the bottom BN layer is exfoliated onto the SiO$_2$ (300 nm)/silicon substrate. The other graphene and BN layers are subsequently transfered using a home-built flake transfer stage with an alignment precision of few micrometers. We employ the flake transfer recipe\cite{taychatanapat2011quantum} based on the sacrificial poly(methyl methacrylate) (PMMA) film coated on supporting polyvinyl alcohol (PVA) layer. After transferring each graphene layer, the device is annealed in H$_2$(5 \%)/Ar(95 \%) at ambient pressure to remove the residual polymers. However PMMA residuals cannot be completely eliminated\cite{lin2011graphene} and exist at the interfaces of graphene and BN. They can still slightly shift the charge neutral point and induce intralayer carrier scatterings where the interlayer scattering from polymer residual may be weak due to large thickness ($\sim$20 nm thick) of the spacer BN layer. Oxygen plasma etching then defines the Hall bar configuration, followed by electron beam lithography and metalization (10 nm Ti and 70 nm Au). The metal lines for the heater and temperature sensor are fabricated together with the metal contacts for the top graphene layer. The device is then covered by the top BN dielectric followed by the top gate metal deposition. The device is finally bonded onto a chip carrier and measured in cryostats. Except the study of temperature dependence, all the other measurements are carried out at room temperature.

The electrical connection to measure the counterflow Seebeck coefficient is shown in Figure \ref{fig:f1} where the counterflow Seebeck coefficient is specifically determined from the voltage $V_\mathrm{S}$ measured between two layers at the right end while they are connected at the left. The lateral temperature difference ($\Delta T$) of $\sim$ 1 K\cite{rmsdeltat} is established between two ends in each graphene layer. The top and bottom gate voltages $V_{\mathrm{tg}}$ and $V_{\mathrm{bg}}$ can control the carrier densities of the corresponding graphene layers, e.g., the top layer of p-type and the bottom layer of n-type here are indicated by the red filling of the parabolic band structures and the charge symbols $\myoplus[gray]$ and $\myominus[red]$.

A low frequency alternating (AC) heating current of a few milliampere in amplitude is applied across the heater line and the counterflow Seebeck voltage $V_\mathrm{S}$ (voltage drop from the top to bottom layer measured at the right ends) is then detected using a lock-in amplifier at the second harmonic (with 90 degree phase shift). The heater and temperature sensors are calibrated\cite{PhysRevLett.91.256801} to calculate the established (RMS) temperature difference $\Delta T$\cite{rmsdeltat} and convert the Seebeck voltage $V_\mathrm{S}$ to Seebeck coefficient $S_{\mathrm{CF}}=V_\mathrm{S}/\Delta T$. The sign of $S_{\mathrm{CF}}$ is negative when the top and bottom graphene layers are p- and n-type respectively. During measuring $V_\mathrm{S}$, the resistances of both graphene layers are simultaneously measured using another two lock-in amplifiers at different lock-in frequencies. The frequencies for the three lock-in amplifiers are varied to ensure that the measured $V_\mathrm{S}$ and resistances are insensitive to such variations, indicating independent thermoelectric and resistance response signals in each graphene layer.

\begin{figure}
	\includegraphics[width=12cm]{f1c.pdf}
	\caption{Schematics of counterflow Seebeck coefficient measurement. The arrows $\myarrowleft[gray]$ and $\myarrowleft[red]$ for the charge carrier symbols $\myoplus[gray]$ and $\myominus[red]$ denote the moving carriers driven by the temperature gradient created by the heater with the velocity $v$ labeled below the arrows. The arrows $\protect\myclosearrowleft$ and $\protect\myclosearrowright$ represent the counterflow electrical currents with magnitude of $I_\mathrm{CF}$ (zero at open circuit). The counterflow Seebeck voltage $V_\mathrm{S}$ is positive where the top (bottom) layer is p-type (n-type).}
	\label{fig:f1}
\end{figure}

\section{Results}

\begin{figure}
  \includegraphics[width=12cm]{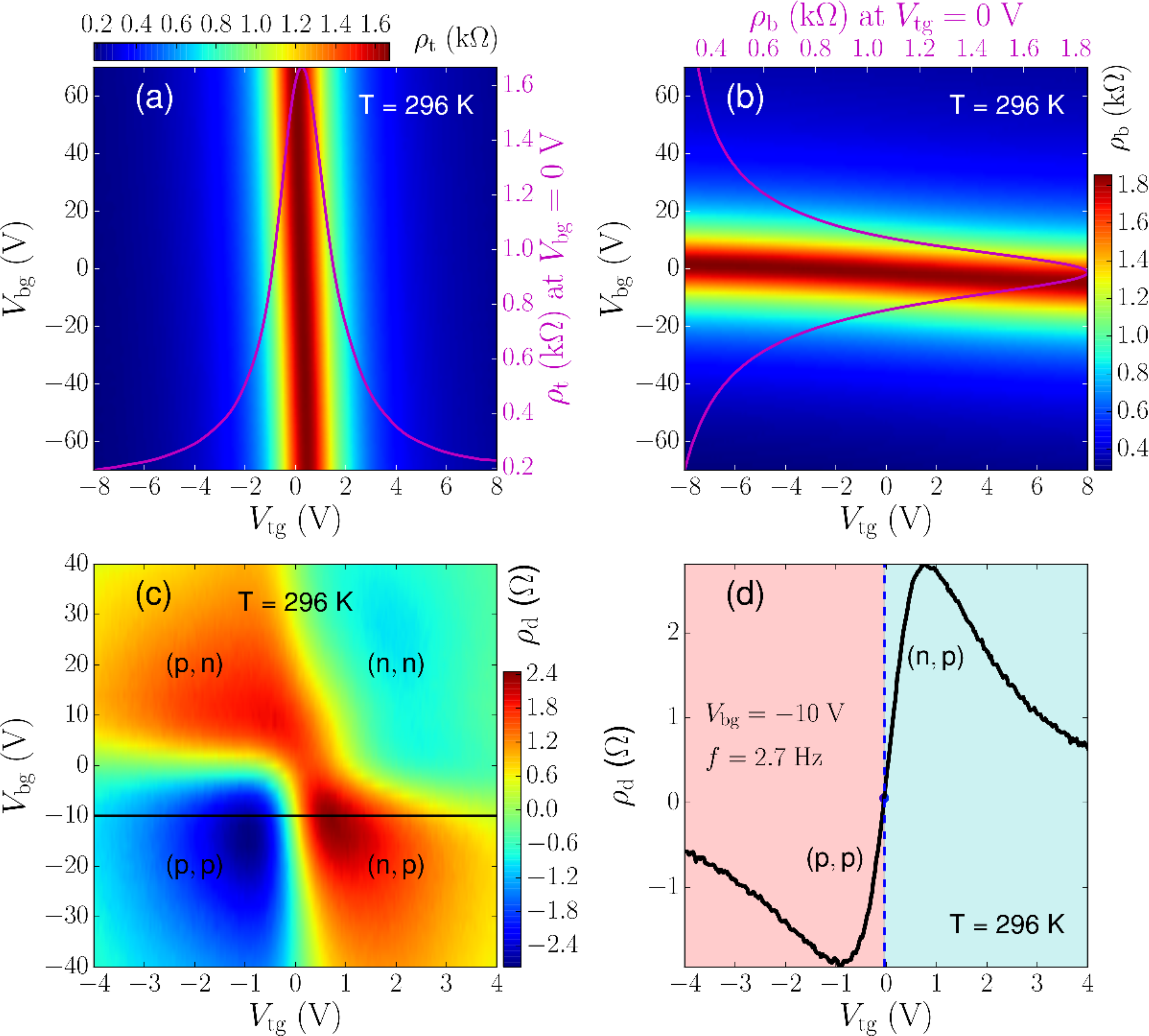}
  \caption{Color maps (a-c) of intralayer resistivities (a) $\rho_\mathrm{t}$ for top BLG, (b) $\rho_\mathrm{b}$ for bottom BLG and Coulomb drag resistivity (c) $\rho_\mathrm{d}$ vs. $V_\mathrm{tg}$ and $V_\mathrm{bg}$, and (d) $\rho_\mathrm{d}$ vs. $V_\mathrm{tg}$ when $V_\mathrm{bg}=-10\;\mathrm{V}$. The purple line in (a) [(b)] is $\rho_\mathrm{t}$ vs. $V_\mathrm{tg}$ [$\rho_\mathrm{b}$ vs. $V_\mathrm{bg}$] when $V_\mathrm{bg} = 0$ [$V_\mathrm{tg} = 0$] with the $\rho_\mathrm{t}$ [$\rho_\mathrm{b}$] plotted at right [top]. Note that (d) is corresponding to the horizontal cut in (c). The vertical blue dashed line in (d) divides it into red and cyan regions indicating p- and n-type of top BLG. The labels (x,y) in (c) and (d) indicate the x[y]-type of charge carriers for top [bottom] BLG, where x and y can be n or p.}
  \label{fig:fetcd}
\end{figure}

The tunneling resistance between two graphene layers is measured to be larger than 10 G$\Omega$, an indication of good electrical isolation between them. The color maps of intralayer resistivities $\rho_\mathrm{t}$ and $\rho_\mathrm{b}$ as functions of $V_\mathrm{bg}$ and $V_\mathrm{tg}$ for the top and bottom BLG are shown in Figure \ref{fig:fetcd}a and b respectively. The solid purple line in Figure \ref{fig:fetcd}a[b] is $\rho_\mathrm{t}$ (plotted on the right axis) vs. $V_\mathrm{tg}$ [$\rho_\mathrm{b}$ (plotted on the top axis) vs. $V_\mathrm{bg}$] when $V_\mathrm{bg}=0$ [$V_\mathrm{tg}=0$]. It is obvious that the resistivity of the top [bottom] BLG layer cannot be effectively tuned by the bottom [top] gate voltage due to the strong screening by the bottom [top] BLG layer. The charge neutral point for the top [bottom] BLG layer slightly decreases as $V_\mathrm{bg}$ [$V_\mathrm{tg}$] increases, due to the incomplete screening from the other graphene layer. The charge neutral points for both graphene layers are very close to zero gate voltage, since both graphene layers are sandwiched between two BN layers, resulting in reduced charge doping of the BLG layers.

The color map of Coulomb drag resistivity $\rho_\mathrm{d}=V_\mathrm{d}/(I_\mathrm{d}N_\mathrm{sq})$ vs. $V_\mathrm{tg}$ and $V_\mathrm{bg}$ is shown in Figure \ref{fig:fetcd}c, where the current $I_\mathrm{d}$ is applied on the drive layer while the voltage $V_\mathrm{d}$ is measured on the drag layer and $N_\mathrm{sq}$ is the ratio of the length along which $I_\mathrm{d}$ is applied to the width of the graphene channel. The sign of $\rho_\mathrm{d}$ is defined to be positive when $V_\mathrm{d}$ has the same sign as the voltage drop in the drive layer when the drag layer has zero current. Various procedures (e.g., swapping the drag and drive layers, changing the grounding positions and varying the lock-in frequencies) are carried out to ensure the physical Coulomb drag signal is detected.\cite{das2011experimental} The Coulomb drag resistivity is measured at a low frequency of $f=2.7\;\mathrm{Hz}$. The out-of-phase component of $\rho_\mathrm{d}$ is at least one order of magnitude smaller than the in-phase component except around Dirac point where $\rho_\mathrm{d}$ is close to zero. The solid line in Figure \ref{fig:fetcd}d is corresponding to the horizontal cut in Figure \ref{fig:fetcd}c at $V_\mathrm{bg}=-10\;\mathrm{V}$. The labels (x,y) in Figure \ref{fig:fetcd}c and d where x and y can be either n or p indicate that the charge carriers of the top BLG is of x-type while that of the bottom BLG is of y-type. Though the shape of the color map is not symmetrical (probably due to charge inhomogeneity), the sign of $\rho_\mathrm{d}$ is correct, i.e, positive for opposite type of charge carriers in both BLG layers (x$\neq$y) and negative for same type of charge carriers (x$=$y). Note that the regions of positive $\rho_\mathrm{d}$ located at the top left and bottom right quadrants are connected across the charge neutral point in both BLG layers, similar to the measurement of Coulomb drag between two single layer graphene (SLG) layers in Ref.\cite{gorbachev2012strong}. For another sample of BLG double layers, the color map of Coulomb drag resistivity (see Figure~S2 in the supplementary material) exhibits different behavior around the charge neutral point in both BLG layers but similar to another reported result for SLG double layers.\cite{kim2012coulomb} The color map of $\rho_\mathrm{d}$ at temperature of 240 K and 200 K can be found in Figure~S1 in the supplementary material. Recent measurements of Coulomb drag in BLG double layers show anomalous negative drag when the carriers in both layers are of the same type at temperatures below $\sim$ 200 K\cite{li2016negative} and 10 K\cite{lee2016giant}. In our sample no such negative drag is observed for temperatures above 200 K. The drag resistivity is too small to allow reliable measurement of Coulomb drag signals.

\begin{figure}
  \includegraphics[width=12cm]{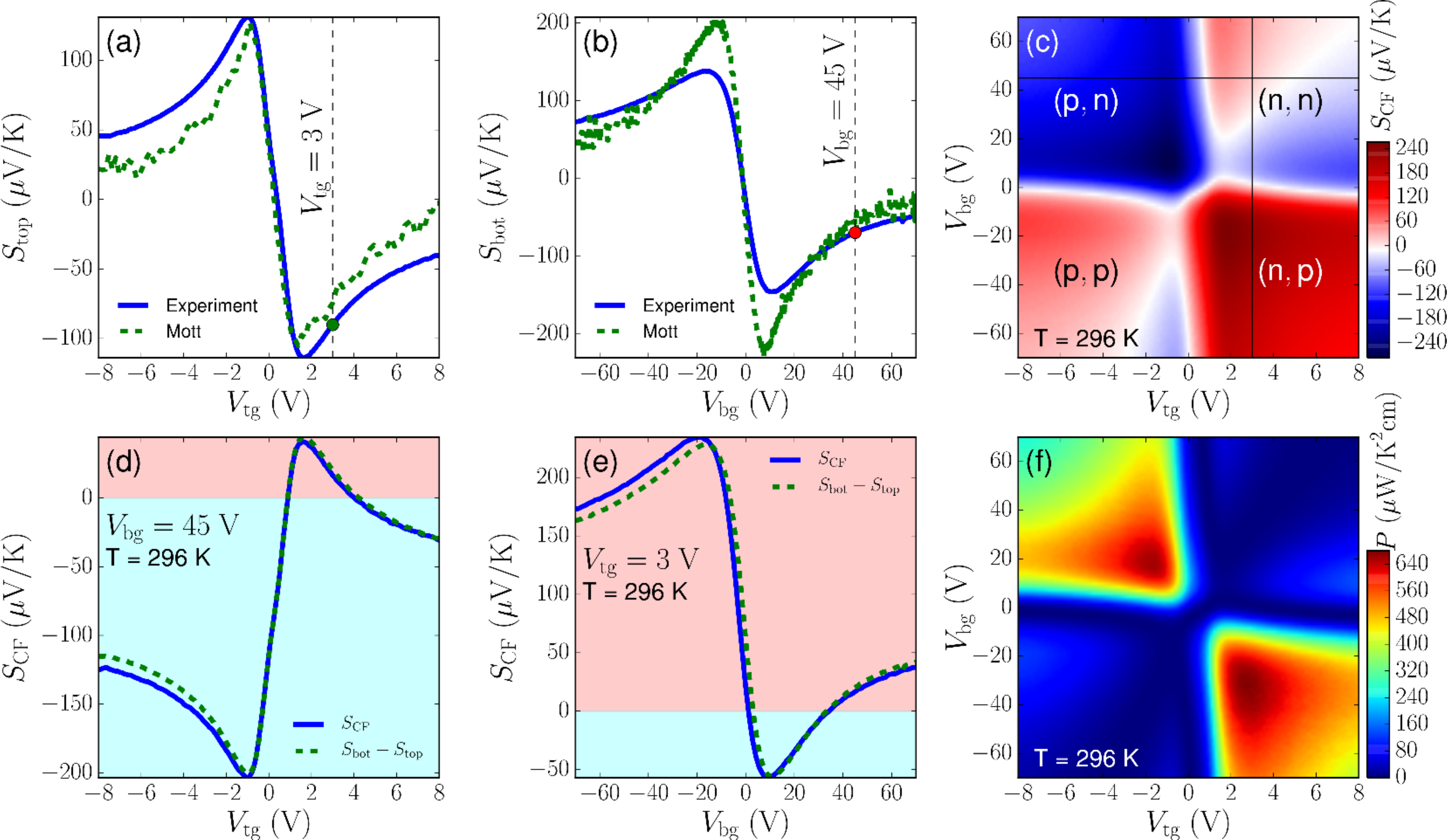}
  \caption{Seebeck coefficient of top (a) and bottom (b) BLG layers (the green dashed lines are calculated Seebeck coefficients from Mott formula) vs. $V_\mathrm{tg}$ and $V_\mathrm{bg}$ respectively, color map (c) of counterflow Seebeck coefficient $S_\mathrm{CF}$ vs. $V_\mathrm{tg}$ and $V_\mathrm{bg}$, (d) $S_\mathrm{CF}$ vs. $V_\mathrm{tg}$ for $V_\mathrm{bg}=45\;\mathrm{V}$, (e) $S_\mathrm{CF}$ vs. $V_\mathrm{bg}$ for $V_\mathrm{tg}=3\;\mathrm{V}$ and color map (f) of power factor $P$ vs. $V_\mathrm{tg}$ and $V_\mathrm{bg}$. The solid blue line in (a) [(b)] is measured within the top [bottom] layer, which is connected to the bottom [top] layer as in Figure~\ref{fig:f1} (there is no observable change in the results if two layers are disconnected).  The solid blue line (measured data) in (d) [(e)] is from the horizontal [vertical] cut in (c). The dashed green line in (d) [(e)] is calculated from $S_\mathrm{bot}-S_\mathrm{top}$ where $S_\mathrm{bot}$ [$S_\mathrm{top}$] is the constant value at $V_\mathrm{bg}=45\;\mathrm{V}$ marked by the red dot in (b) [at $V_\mathrm{tg}=3\;\mathrm{V}$ marked by the green dot in (a)] while $S_\mathrm{top}$ [$S_\mathrm{bot}$] is the solid blue line in (a) [(b)]. The red [cyan] area in (d) and (e) represents the positive [negative] value of $S_\mathrm{CF}$.}
  \label{fig:scf}
\end{figure}

The blue solid lines in Figure \ref{fig:scf}a[b] are the measured Seebeck coefficient $S_{\mathrm{top[bot]}}$ of the top [bottom] BLG layer tuned by the top [bottom] gate voltage $V_{\mathrm{tg[bg]}}$ while the bottom [top] BLG layer is electrically floating. The green dashed lines are calculated from the Mott formula $S=\frac{\pi^2k_\mathrm{B}^2T}{3e\rho}\frac{\mathrm{d}\rho}{\mathrm{d}V_\mathrm{g}}\frac{\mathrm{d}V_\mathrm{g}}{\mathrm{d}E_\mathrm{F}}$, where $e$ is the elementary charge, $k_\mathrm{B}$ is the Boltzmann constant, $V_\mathrm{g}$ is the gate voltage, $\rho$ is the intralayer resistivity, and $E_\mathrm{F}$ is the Fermi level.\cite{nam2010thermoelectric} The $V_\mathrm{g}$ vs. $E_\mathrm{F}$ relation can be obtained by evaluating the energy dispersion $E(k)=\pm\frac{1}{2}\gamma_1\left[\sqrt{1+4v_\mathrm{F}^2\hbar^2k^2/\gamma_1^2}-1\right]$ ($+$ for conduction band and $-$ for valence band) of BLG at the Fermi wave vector $k_\mathrm{F}=\sqrt{\pi C_\mathrm{g}|V_\mathrm{g}-V_\mathrm{D}|/e}$, where the Fermi velocity is $v_\mathrm{F}=10^6$ m/s, $\gamma_1=0.39$ eV, $C_\mathrm{g}$ is the gate capacitance and $V_\mathrm{D}$ is the gate voltage at the charge neutral point.\cite{mccann2006landau} Note that the wiggles in the green dashed lines are due to the numerical computation of $\frac{\mathrm{d}\rho}{\mathrm{d}V_\mathrm{g}}$ from the solid purple lines in Figure \ref{fig:fetcd}a and b.

The color map of counterflow Seebeck coefficient $S_{\mathrm{CF}}$ ($=V_\mathrm{S}/\Delta T$) vs. $V_{\mathrm{tg}}$ and $V_{\mathrm{bg}}$ at room temperature is shown in Figure \ref{fig:scf}c (temperature dependence of $S_{\mathrm{CF}}$ can be found in Figure~S3 in the supplementary material). Four quadrants of this color map are separated by white areas near zero gate voltages. Large magnitude of negative [positive] $S_{\mathrm{CF}}$ appears in the top left [bottom right] quadrant when the top and bottom graphene layers are p- and n-type [n- and p-type] respectively. In this regime where two graphene layers have different carrier types, we found that the magnitude of $S_{\mathrm{CF}}$ is simply the sum of the magnitude of Seebeck coefficient in each individual layer. We do not find any noticeable interlayer interaction effects on $S_{\mathrm{CF}}$ due to the weak Coulomb drag in our samples ($\rho_\mathrm{d}$ is nearly three orders of magnitude smaller than $\rho_\mathrm{t}$ or $\rho_\mathrm{b}$). The magnitude of $S_{\mathrm{CF}}$ in the other two quadrants is smaller or even close to zero since the signs of Seebeck coefficient in both layers are same and they tend to cancel each other. The above discussion can be represented by the formula $S_\mathrm{CF}=S_\mathrm{bot}-S_\mathrm{top}$ which is demonstrated in Figure \ref{fig:scf}d-e. For example in Figure \ref{fig:scf}d, the solid blue line is from the horizontal cut in Figure \ref{fig:scf}c while the dashed green line is calculated from $S_\mathrm{bot}-S_\mathrm{top}$ where $S_\mathrm{bot}$ is the constant value taken at the red point in Figure \ref{fig:scf}b for $V_\mathrm{bg}=45\mathrm{V}$ and $S_\mathrm{top}$ is the solid blue curve in Figure \ref{fig:scf}a. These two lines are close to each other, validating the above formula. The small discrepancy may come from the incomplete gating screening of BLG layers, resulting in small variation of the carrier density of the top [bottom] BLG when $V_\mathrm{bg}$ [$V_\mathrm{tg}$] is changed and $V_\mathrm{tg}$ [$V_\mathrm{bg}$] is fixed. The horizontal lines of zero $S_\mathrm{CF}$ in Figure \ref{fig:scf}d-e separate them into red (positive $S_\mathrm{CF}$) and cyan (negative $S_\mathrm{CF}$) areas. Note that the lines of $S_\mathrm{CF}$ vs. gate voltage cross zero twice, which is not possible for an isolated and homogeneous single layer of BLG. This explains that the top right and bottom left quadrants in Figure \ref{fig:scf}c are composed of regions with different sign of $S_{\mathrm{CF}}$ and the boundaries of white line segments with zero $S_{\mathrm{CF}}$.

The power factor in the counterflow thermoelectric transport regime is defined as $P=S_{\mathrm{CF}}^2/[(\rho_\mathrm{t}+\rho_\mathrm{b})t]$, where the BLG thickness $t$ (=0.67 nm) is introduced to allow convenient comparison of our measured results with the other experiments and calculations. Its color map is shown in Figure \ref{fig:scf}f. Note that the gate voltages corresponding to the maximum of $P$ are different from that of the magnitude of $S_{\mathrm{CF}}$ since the BLG resistivity is strongly dependent on the gate voltages. Here the extrinsic resistance mainly from the metal-graphene contacts is ignored, because it is material and process dependent and can be one order smaller than the graphene resistances after special contact treatment.\cite{leong2013low} The maximum power factor is $P_\mathrm{max}\sim$ 700 $\mu$W/K$^2$cm. A very recent work has obtained similar power factor in graphene/BN devices.\cite{duan2016high}

\begin{figure}
  \includegraphics[width=12cm]{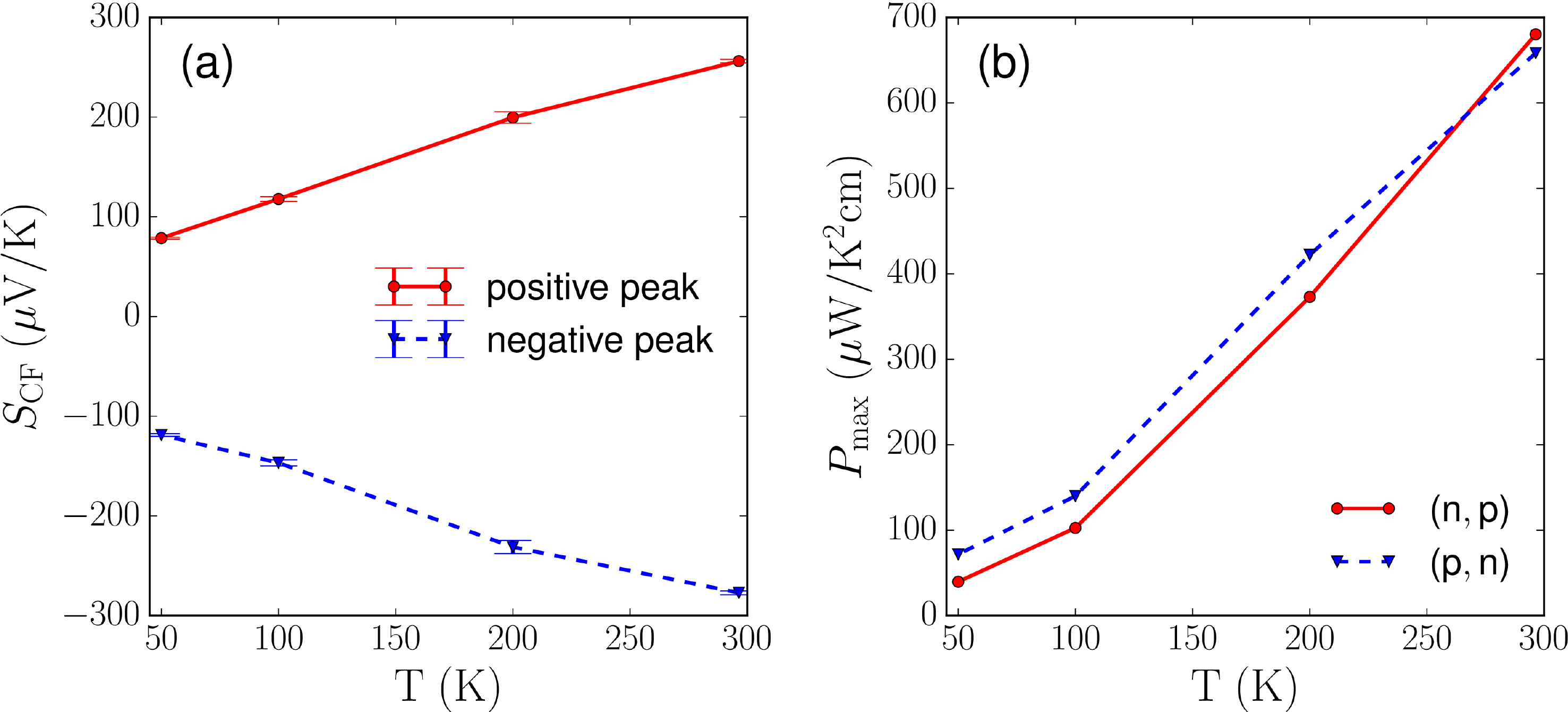}
  \caption{Positive and negative peaks of $S_\mathrm{CF}$ (a) and the maximum power factor (b) $P_\mathrm{max}$ vs. temperature. The error bars (type A standard uncertainties\cite{bipm2008evaluation}) in (a) mainly come from the calibration for the temperature sensors. The positive/negative peaks in $S_\mathrm{CF}$ and positive peaks in $P$ are taken from the corresponding color maps showing their dependence on $V_\mathrm{tg}$ and $V_\mathrm{bg}$ at various temperatures.}
  \label{fig:scftdep}
\end{figure}

The positive and negative peak values of $S_\mathrm{CF}$ vs. temperature are shown in Figure \ref{fig:scftdep}a. The magnitudes of both peaks decrease with temperature. 
The maximum power factor $P_\mathrm{max}$ in the two regimes of opposite carrier types for the two graphene layers (solid red [dashed blue] line for top layer of n-type [p-type] and bottom layer of p-type [n-type]) vs. temperature is shown in Figure \ref{fig:scftdep}b. These two lines show similar trend and manifest the symmetry between the two regimes. It is expected that $P_\mathrm{max}$ will increase further as temperature increases above room temperature. Note that the peak values of $S_\mathrm{CF}$ and $P_\mathrm{max}$ are both achieved in $(\mathrm{p},\mathrm{n})$ and $(\mathrm{n},\mathrm{p})$ regions, but the corresponding gate voltages are different.

\section{Discussion}

Although graphene has high Seebeck coefficient and electrical conductivity,\cite{chen2008intrinsic,sankeshwar2013thermoelectric} hence high power factor, its $ZT$ is small due to its high thermal conductivity.\cite{balandin2011thermal,pop2012thermal} While small $ZT$ materials are not good candidates for thermoelectric applications, significant effort has been devoted to enhancing the $ZT$ of graphene and related nanostructures,\cite{sevinccli2010enhanced,chen2010thermoelectric,mazzamuto2011enhanced,huang2011theoretical,gunst2011thermoelectric,liang2012enhanced,zheng2012enhanced,xie2012enhancement,kliros2012thermoelectric,yang2012enhanced,karamitaheri2012engineering,chang2012edge,pan2012ballistic,sevinccli2013bottom,yeo2013first,feng2013enhanced,chen2013thermoelectric} usually by reducing the thermal conductivity with various methods. The high value of power factor is still promising in electricity generation at special circumstances, e.g., when space is constrained and high efficiency is not needed. The maximum output power for the thermoelectric module with $N_T$ serially connected (see Figure~S5 in the supplementary material) units is $Q_\mathrm{m}=P_\mathrm{max} N_Tt\Delta T^2/4N_\mathrm{sq}$ at the impedance match\cite{apertet2012optimal} where the load resistance equals to the internal resistance ($=N_T(\rho_\mathrm{t}+\rho_\mathrm{b})N_\mathrm{sq}$) of the module, where $N_\mathrm{sq}$ is the ratio of the length $L$ (the length of graphene is measured along the direction of temperature gradient) to the width $W$ and $P_\mathrm{max}\sim$ 700 $\mu$W/K$^2$cm. The expression of $Q_\mathrm{m}$ indicates that the effective power factor ($=N_TP_\mathrm{max}$) of $N_T$ serially connected units is proportional to $N_T$, useful for building high output voltage thermoelectric modules. For large $N_T$, the internal resistance of the module will be proportionally large and the power delivery capability will be limited when the load resistance is a constant. This issue can be resolved by reducing $N_\mathrm{sq}$. The values of $N_T$ and $N_\mathrm{sq}$ can be optimally chosen for any specific load resistance and power requirement.

For $\Delta T=50$ K (e.g., the average temperature difference between the hot and cold spots in the modern CPUs), $Q_\mathrm{m}=2.9$ W for $1/N_\mathrm{sq}=N_T=10000$. Such serially connected thermopower generator has a thermal conductance of $G=2\kappa N_Tt/N_\mathrm{sq}=201$ W/K for $\kappa=1500$ W/m$\cdot$K ($G$ will be larger if the heat conduction channels through the insulating materials between graphene layers and the other supporting/gating materials are included), useful for applications of fast heat dissipation while generating electricity energy. Due to the low mass density of graphene, the power output per mass is as high as $\sim\;1.1\times 10^9$ W/kg (the total graphene area is $\sim$ 20 $\mu$m$^2$) without including the mass of the insulating materials between graphene layers. It is still an extremely large number even when the mass of those insulating materials are included. For example, inserting a 3 nm thick of BN layer between graphene layers is good enough to avoid leakage between graphene layers. The mass density of BN is about 4.5 times larger than that of graphene, thus the power output per mass is reduced to $\sim\;2.4\times 10^8$ W/kg. The mass of the other materials such as the gate metal (can be replaced by graphene or become not needed by doping graphene chemically for example) and interconnects (could be replaced by graphene) is not considered here. Such high power per mass makes graphene promising in light weight thermoelectric applications.\cite{yazawa2011cost}

The intralayer resistivities may be affected by the interlayer Coulomb drag,\cite{tutuc2004counterflow} which will affect the power factor. As seen from Figure \ref{fig:fetcd}c, the maximum magnitude of $\rho_\mathrm{d}$ is about 2.5 $\Omega$, much smaller than the resistivity of each graphene layer. Therefore, such Coulomb drag effect can be neglected in the counterflow thermoelectric transport measured here. It may become important when $\rho_\mathrm{d}$ is large, especially when it is close to the intralayer resistivities. Large Coulomb drag resistivity is an indication of strong interlayer interactions. It has been suggested that the formation of excitons composed of electron-hole pairs where the electrons and holes reside in different layers will possibly lead to high $ZT$ structures.\cite{wu2014bilayer,chen2012surface,huexciton2017}

\section{Conclusion}

In conclusion, Coulomb drag and counterflow thermoelectric transport measurements are performed in layered structures of BN/BLG/BN/BLG/BN. The magnitude of the counterflow Seebeck coefficient exhibits a peak in the regime where two graphene layers have opposite sign of charge carriers. The maximum power factor is about $700\mu$W/K$^2$cm at room temperature. A quantitative analysis indicates that graphene can be useful for light weight thermoelectric systems. The counterflow Seebeck coefficient and power factor decrease with temperature from 300 K to 50 K. The measured interlayer Coulomb drag resistivity is small ($<$ 3 $\Omega$) compared to the intralayer resistivity, making negligible impact of Coulomb drag on the counterflow thermoelectric transport in the present structures. However, further reducing the separation between two BLG layers may shed light on the interplay between Coulomb drag and counterflow thermoelectric transport.

\section*{Acknowledgment}
This work is supported by DARPA MESO (grant N66001-11-1-4107) and NIST MSE (grant 60NANB9D9175).

\section*{References}

\begin{thebibliography}{10}
\bibitem{rmsdeltat}
Here 1 K is the RMS value of the temperature difference $\Delta T$ that is
  oscillating with a frequency of $2\omega$ when an AC current with frequency
  $\omega$ is applied to the heater. $\Delta T$ is extracted from the measured
  $2\omega$ voltages with 90$^\circ$ phase shift on the sensors.
\end{thebibliography}
\providecommand{\noopsort}[1]{}\providecommand{\singleletter}[1]{#1}%

\section*{Supporting Information}

The color maps of Coulomb drag resistivity $\rho_\mathrm{d}$ at T = 240 K and T = 200 K are shown in Figure \ref{fig:cdtdep}a and b respectively. The data below 200 K is not collected due to small signal to noise ratio of the Coulomb drag signal. The measured data exhibits the anomalous behavior that the magnitude of the red peak of $\rho_\mathrm{d}$ increases when the temperature is reduced. This is possibly due to the sample inhomogeneity. 

\renewcommand{\thefigure}{S\arabic{figure}}
\setcounter{figure}{0}

\begin{figure}[!htb]
  \includegraphics[width=10cm]{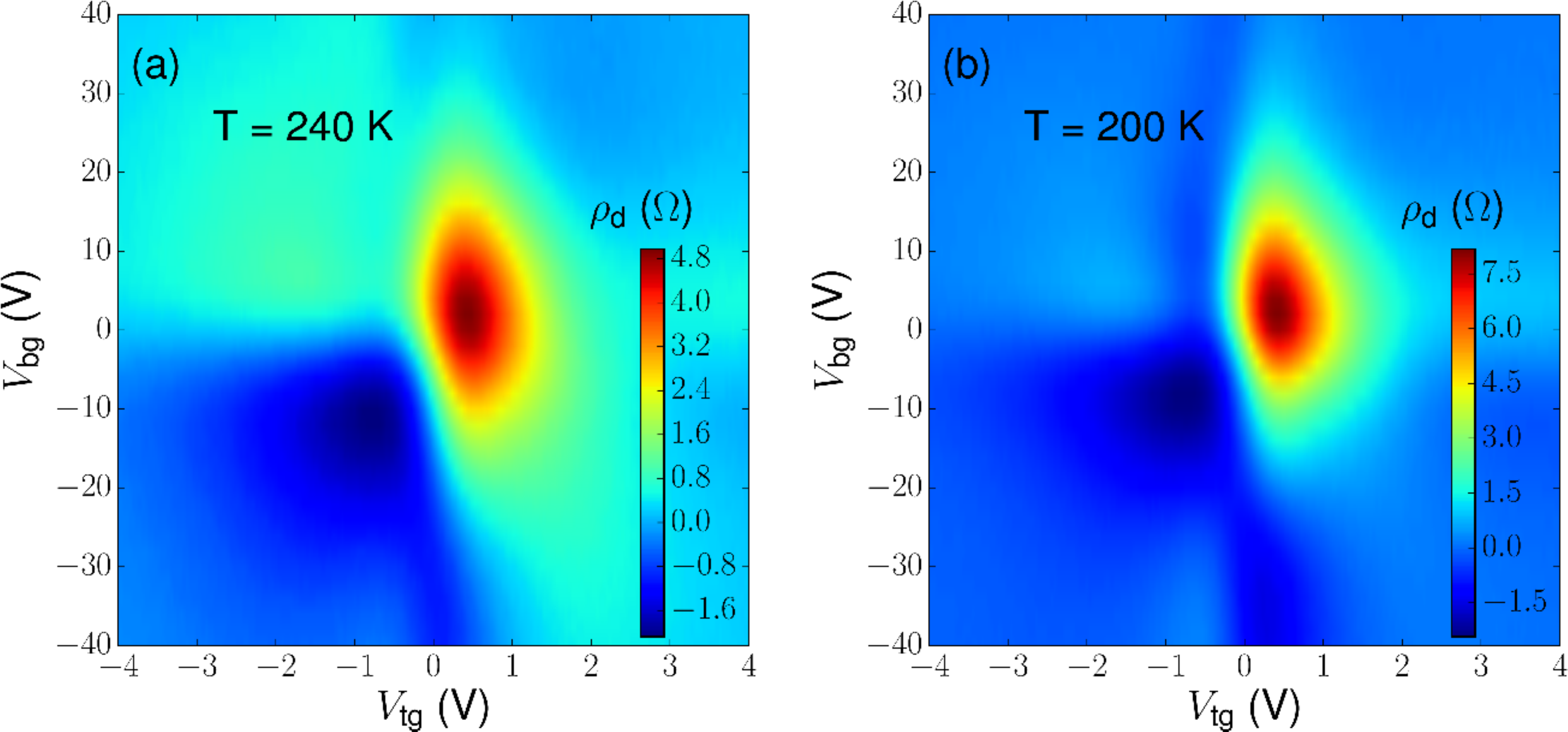}
  \caption{(a) $\rho_\mathrm{d}$ at T = 240 K, (b) $\rho_\mathrm{d}$ at T = 200 K.}
  \label{fig:cdtdep}
\end{figure}

\begin{figure}[!htb]
  \includegraphics[width=10cm]{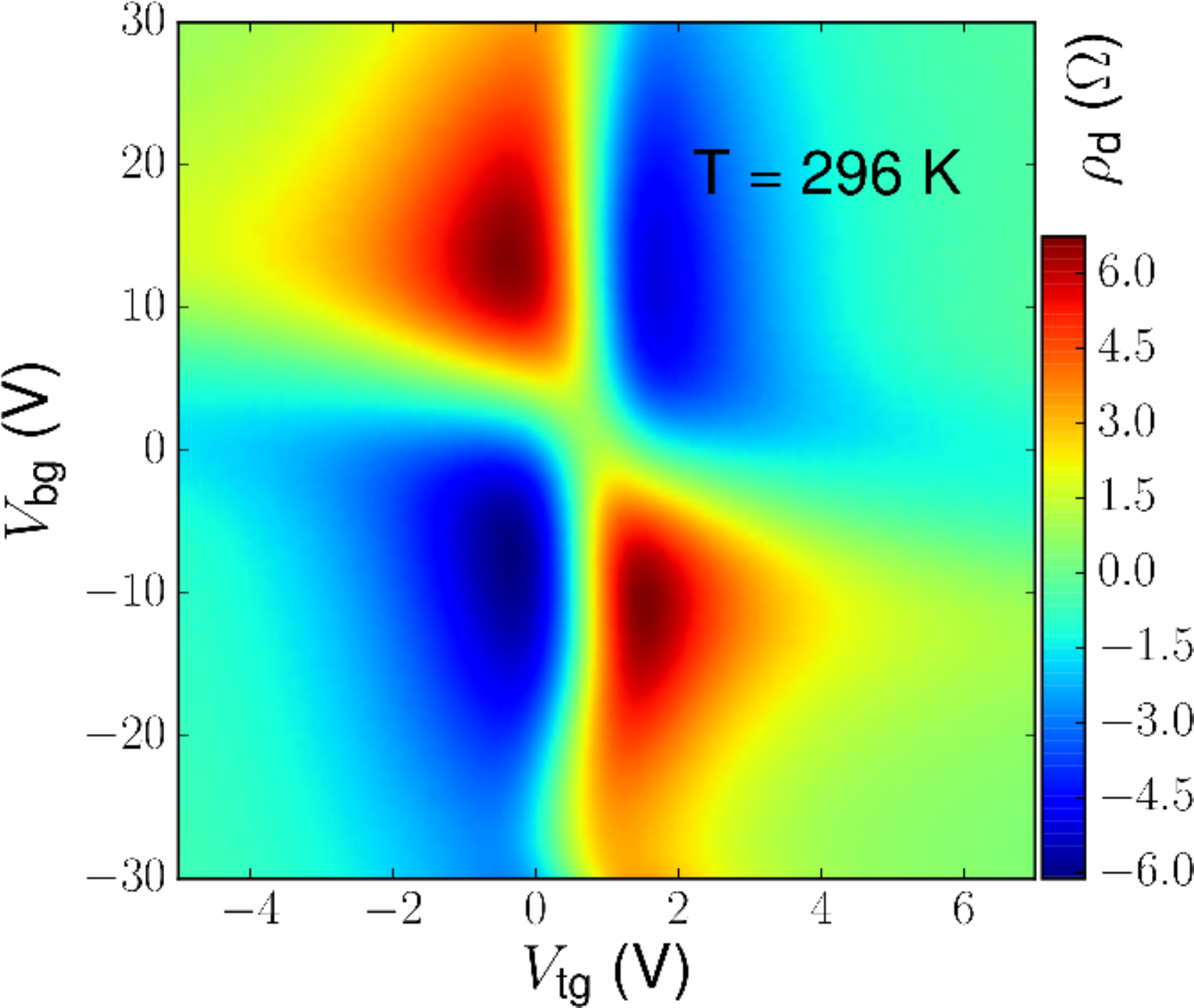}
  \caption{Color map of Coulomb drag resistivity $\rho_\mathrm{d}$ vs. $V_\mathrm{tg}$ and $V_\mathrm{bg}$ for another sample of bilayer-graphene double layers measured at room temperature.}
  \label{fig:cdhbn146}
\end{figure}

\begin{figure}[!htb]
  \includegraphics[width=10cm]{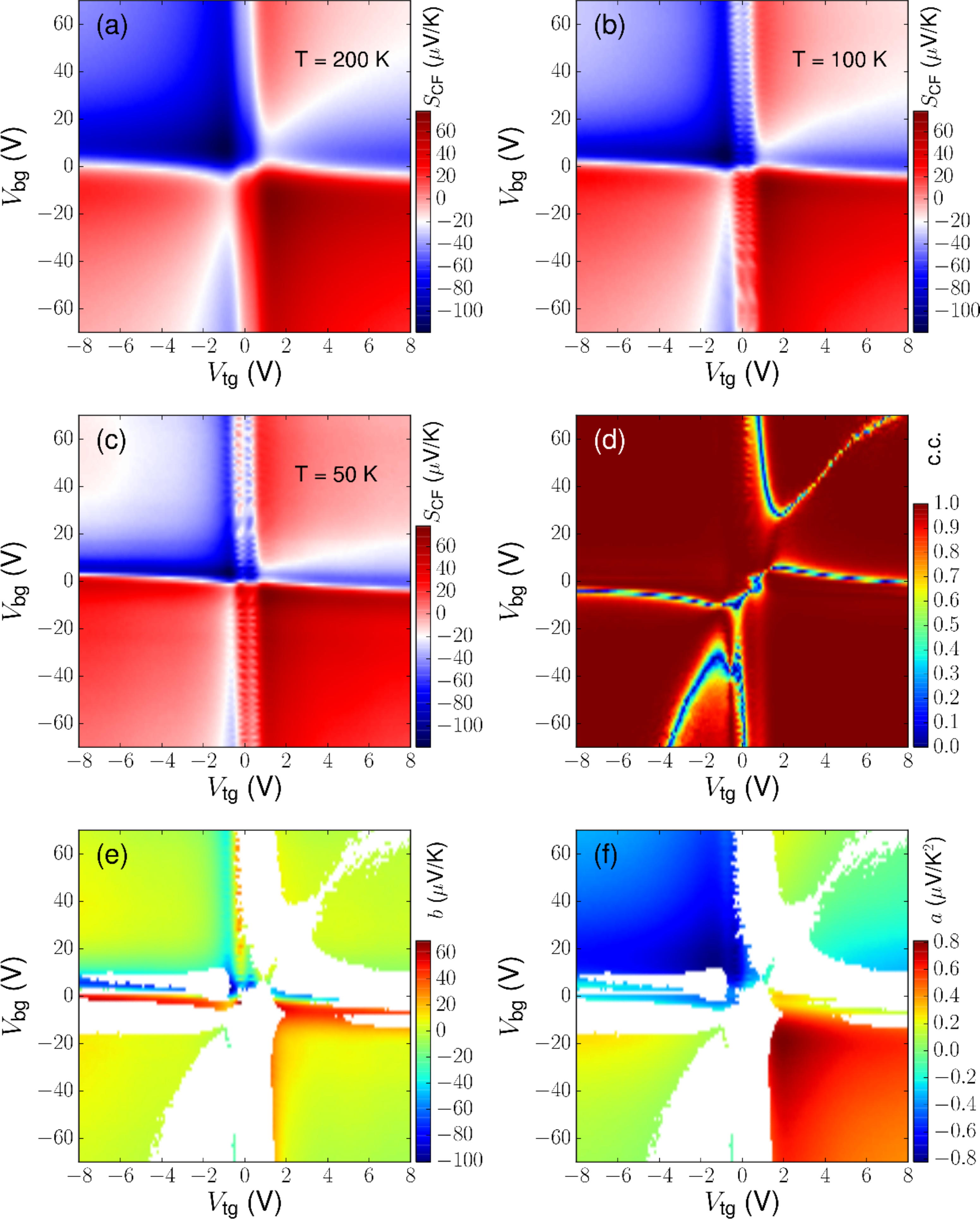}
  \caption{Color maps (a, b and c) of $S_\mathrm{CF}$ vs. $V_\mathrm{tg}$ and $V_\mathrm{bg}$ at T = 200 K, 100 K and 50 K respectively, (d) correlation coefficient (c.c.) between $S_\mathrm{CF}$ and temperature, color maps of the intercept $b$ (e) and slope $a$ (f) of the linear fit of $S_\mathrm{CF}\sim a\mathrm{T}+b$. The blank area in (e) and (f) corresponds to the gate voltages at which c.c. in (d) is less than 0.99.}
  \label{fig:scf100k}
\end{figure}

\begin{figure}[!htb]
  \includegraphics[width=10cm]{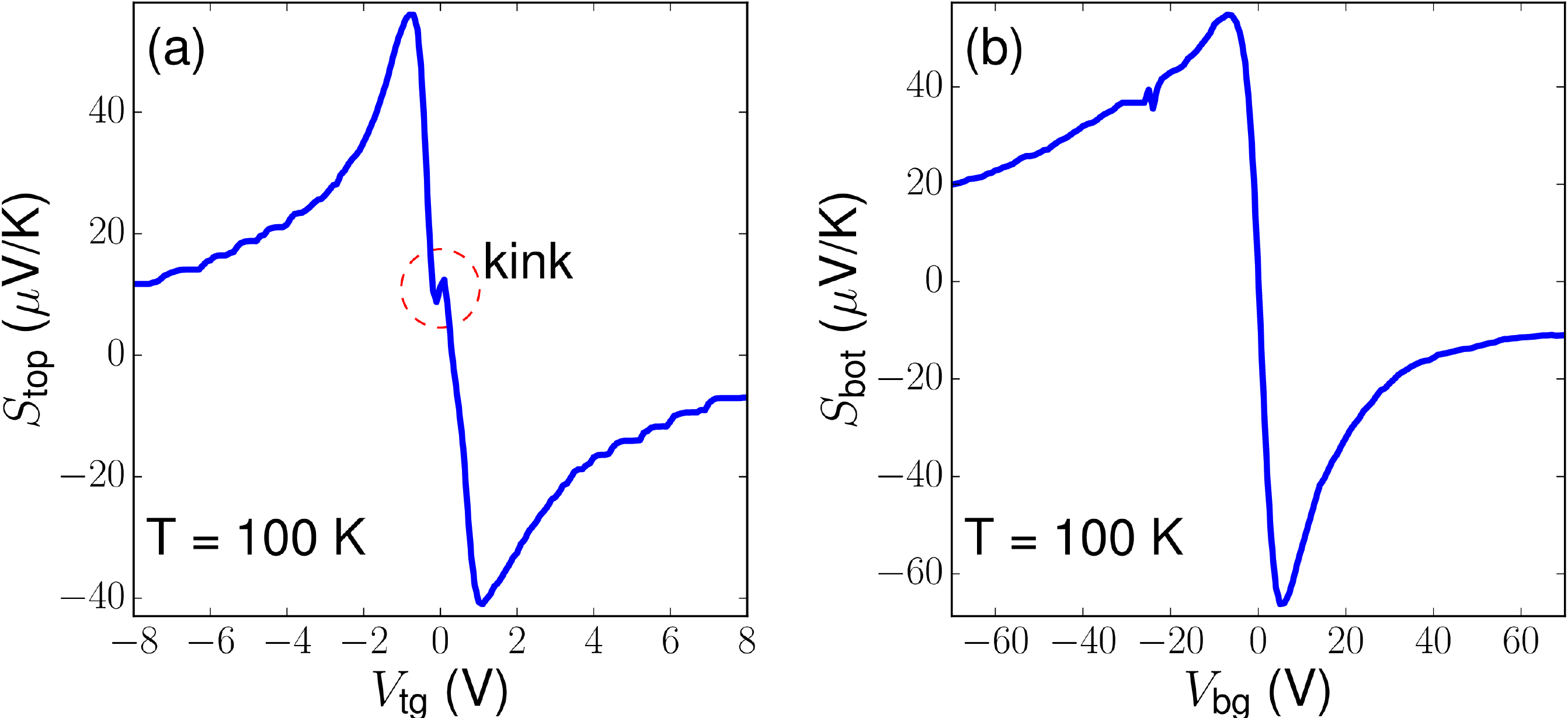}
  \caption{Seebeck coefficient for top (a) and bottom (b) BLG layers. The kink in (a) inside the red dashed circle is labeled.}
  \label{fig:scfkink}
\end{figure}

The color maps of $S_\mathrm{CF}$ at T = 200 K, 100K and 50 K are shown in Figure \ref{fig:scf100k}a-c. As the temperature decreases, $S_\mathrm{CF}$ decreases and the red area of positive $S_\mathrm{CF}$ increases for the top right and bottom left quadrants in the color maps. The positive and negative peaks of $S_\mathrm{CF}$ are slightly shifted closer to zero gate voltages and become more concentrated as the temperature is reduced. Many puddles of $S_\mathrm{CF}$ appear mainly around $V_\mathrm{tg}=$0 V for temperatures 100 K and 50 K at which the Seebeck coefficient of the top BLG vs. $V_\mathrm{tg}$ exhibits kinks (see Figure~\ref{fig:scfkink}) around the charge neutral point of top BLG, unlike that of the bottom BLG smoothly crosses zero. This could be due to the manifestation of the charge inhomogeneity in the top BLG layer at lower temperatures and possibly the small misalignment of the two BLG layers during sample fabrication, resulting in disturbed Seebeck coefficient of the top BLG layer around its charge neutral point. The color maps of the correlation coefficient (c.c., defined later) of $S_\mathrm{CF}$ vs. T and the intercept $b$ and the slope $a$ for the relation $S_\mathrm{CF}\sim a\mathrm{T}+b$ are shown in Figure \ref{fig:scf100k}d, e and f respectively. The blank area in Figure \ref{fig:scf100k}d and f corresponds to the gate voltages at which the c.c. in Figure \ref{fig:scf100k}d is less than 0.99. The magnitude of the intercept in Figure \ref{fig:scf100k}e is close to zero for the majority area of gate voltages away from the charge neutral point, indicating that $S_\mathrm{CF}$ is indeed linearly dependent on temperature.

The c.c. between two sets of data, e.g., \{$x_1,x_2,\cdots,x_n$\} and \{$y_1,y_2,\cdots,y_n$\} is defined as $\mathrm{c.c.}=\frac{\sum_{i=1}^n|(x_i-\bar{x})(y_i-\bar{y})|}{\sqrt{\sum_{i=1}^n(x_i-\bar{x})^2\sum_{i=1}^n(y_i-\bar{y})^2}}$ where $\bar{x}=\frac{\sum_{i=1}^nx_i}{n}$ and $\bar{y}=\frac{\sum_{i=1}^ny_i}{n}$. The magnitude of c.c. measures the linearity between the two data sets: c.c. = 1 implies that the two sets are perfectly linearly dependent.\cite{jaynes2003prob}

\begin{figure}[!htb]
  \includegraphics[width=10cm]{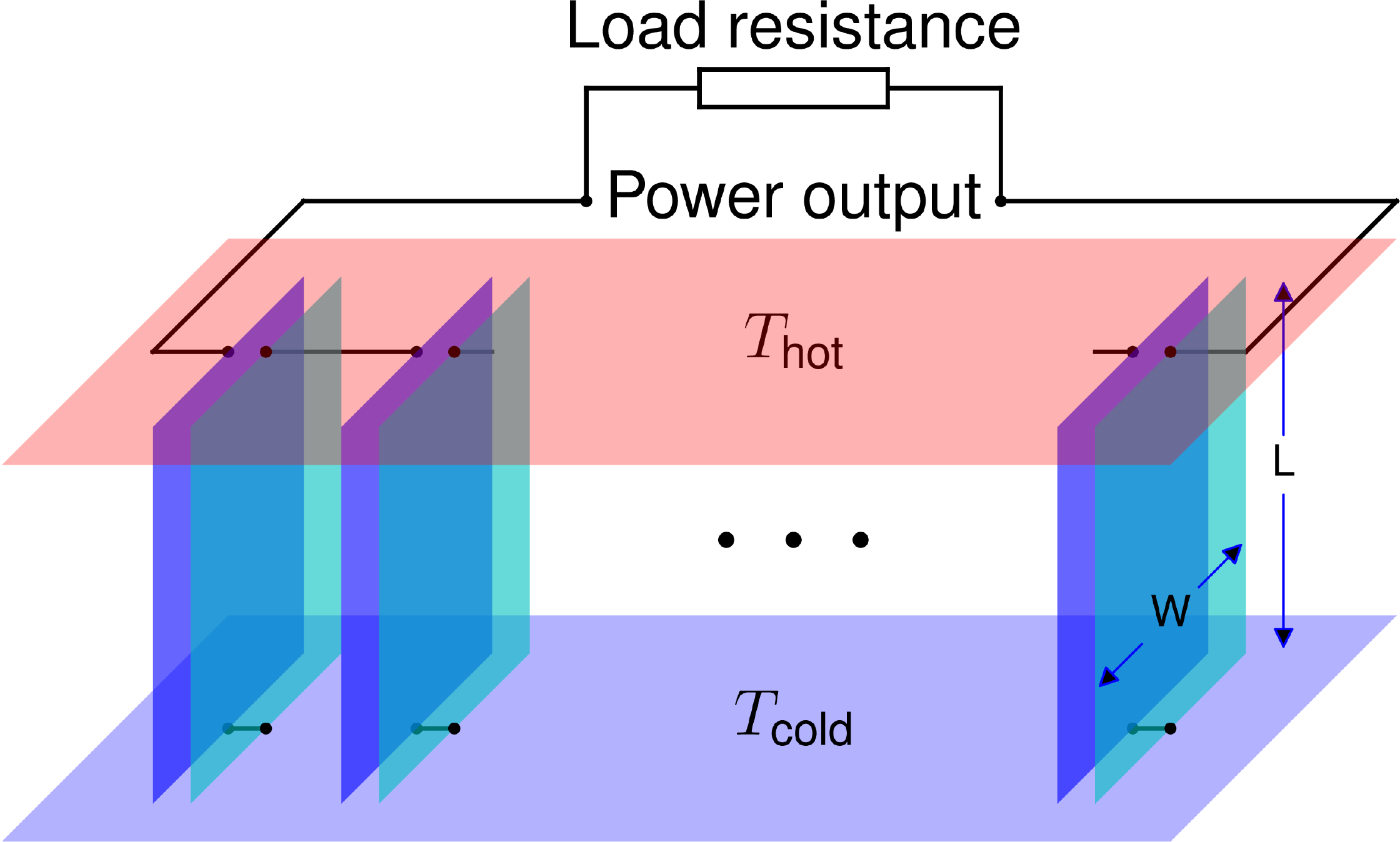}
  \caption{Multiple double layer (vertical blue for n-type and cyan for p-type) units connected in serial. The top and bottom ends of the units are placed at hot and cold reservoirs with temperatures $T_{\mathrm{hot}}$ and $T_{\mathrm{cold}}$ respectively. The channel width (W) and length (L) for one unit is labeled. The gate metal and BN spacers are not shown. The load resistance is connected at the two ends of the power output.}
  \label{fig:cfconnection}
\end{figure}

\end{document}